\documentclass[a4paper,11pt]{article}
\pdfoutput=1 

\usepackage{jcappub}
\usepackage[T1]{fontenc}
\usepackage{amssymb,amsmath,amsfonts,amsbsy,graphicx,microtype,rotating,bm}

\renewcommand{\(}{\left(}
\renewcommand{\)}{\right)}
\renewcommand{\[}{\left[}
\renewcommand{\]}{\right]}

\newcommand{\mpl}{m_{\rm Pl}}
\newcommand{\fnl}{f_{\rm NL}}
\newcommand{\calO}{\mathcal{O}}
\newcommand{\calP}{\mathcal{P}}
\newcommand{\calR}{\mathcal{R}}

\usepackage{color}
\newcommand{\tc}{\textcolor{blue}}

\subheader{APCTP-Pre2016-010}

\title{\boldmath Features from the non-attractor beginning of inflation}

\author[a]{Yi-Fu Cai,}
\author[b,c]{Jinn-Ouk Gong,}
\author[a]{Dong-Gang Wang,}
\author[a]{and Ziwei Wang}

\affiliation[a]{CAS Key Laboratory for Researches in Galaxies and Cosmology,  University of Science and Technology of China, Chinese Academy of Sciences, Hefei, Anhui 230026, China}
\affiliation[b]{Asia Pacific Center for Theoretical Physics, Pohang 37673, Korea}
\affiliation[c]{Department of Physics, Postech, Pohang 37673, Korea}

\emailAdd{yifucai@ustc.edu.cn}
\emailAdd{jinn-ouk.gong@apctp.org}
\emailAdd{wdgang@mail.ustc.edu.cn}
\emailAdd{wzw2013@mail.ustc.edu.cn}

\abstract{
We study the effects of the non-attractor initial conditions for the canonical single-field inflation. The non-attractor stage can last only several $e$-folding numbers, and should be followed by hilltop inflation. This two-stage evolution leads to large scale suppression in the primordial power spectrum, which is favored by recent observations. Moreover we give a detailed calculation of primordial non-Guassianity due to the ``from non-attractor to slow-roll'' transition, and find step features in the local and equilateral shapes. We conclude that a plateau-like inflaton potential with an initial non-attractor phase yields interesting features in both power spectrum and bispectrum.}


\arxivnumber{1607.07872}

\begin{document}

\maketitle
\flushbottom

\section{Introduction}
\label{intro}

Inflationary scenario is the most popular paradigm for the very early universe which successfully resolves the puzzles of standard big bang cosmology~\cite{Starobinsky:1980te,Fang:1980wi,Guth:1980zm,Sato:1980yn,Linde:1981mu, Albrecht:1982wi}. Conventionally, inflation is driven by the potential of a slowly rolling canonical scalar field, the inflaton, and this evolution is described by the smallness of the so-called slow-roll parameters. It predicts a nearly scale-invariant power spectrum of the primordial curvature perturbation, which has been verified to high precision by the latest measurements of cosmic microwave background (CMB) temperature anisotropies~\cite{Ade:2015xua,Ade:2015lrj}.

From the observational aspects, however, there are several issues remaining to be settled. For example, the primordial non-Gaussianity, as a powerful tool to distinguish among various models of inflation and alternatives, has not been detected. Also the recent CMB observations seem to indicate a relative power suppression on low multipoles~\cite{Ade:2015lrj}, which deviates from the prediction of naive single-field slow-roll inflation. These unresolved anomalies may well indicate that, upon further supply of more accurate observational data, our fiducial model of canonical single-field slow-roll inflation is not sufficient, and that we indeed observe the trail of the earlier phase of inflation or its alternatives.

One theoretical direction to accommodate the observed CMB anomalies is to abandon the {\em assumption} of slow-roll, attractor evolution of the inflaton from beginning to end. Indeed, a few models of non-attractor phase of inflation have been suggested and studied, with one representative model being the ultra-slow-roll inflation~\cite{Tsamis:2003px,Kinney:2005vj}. In this model, the first slow-roll parameter remains very small but the second one is $\mathcal{O}(1)$ during inflation, leading to the growth of the curvature perturbation after horizon exit rather than remaining constant (see also~\cite{Motohashi:2014ppa}). Then it is realized that the non-attractor evolution can lead to a large bispectrum in the squeezed configuration even in single-field inflation models with a Bunch-Davies vacuum~\cite{Namjoo:2012aa,Martin:2012pe,Chen:2013aj,Chen:2013eea}, which violates the non-Gaussianity consistency relation~\cite{Maldacena:2002vr,Creminelli:2004yq}. This behavior of the curvature perturbation is the same as the one in the matter bounce cosmology, where the matter dominated contraction can also be seen as a ``non-attractor'' stage~\cite{Cai:2009fn, Cai:2014bea, Brandenberger:2016vhg}.

Moreover, the latest CMB observations give rise to an upper bound for the tensor-to-scalar ratio as $r<0.07$~\cite{Ade:2015lrj, Array:2015xqh}. This result favors the inflation models with a very flat, plateau-like potential, such as the $R^2$ inflation~\cite{Starobinsky:1980te} and the recent $\alpha$-attractors~\cite{Kallosh:2013yoa}. There have been more and more discussions about the plateau-like potentials constructed from fundamental theories and their conceptual issues, such as eternal inflation~\cite{Ijjas:2015hcc}. The slow-roll trajectories of these models usually begin with a nearly vanishing field momentum at large field values. However, this needs not necessarily be the case and an initial non-attractor stage with relatively large field velocity can provide preferred observational consequences. For example, a preceding fast-roll stage on a steeper fraction of the potential may well provide a large field velocity~\cite{Jain:2008dw}. Therefore it is interesting to study the effects of these more general initial conditions.

In this work, we phenomenologically study the non-attractor beginning of inflation on the plateau-like potential. We first note that the duration of canonical non-attractor inflation is limited to several $e$-folding numbers, and a hilltop potential with slow-roll evolution is needed to render it complete. It is shown that because of the relaxed initial condition, the primordial spectrum is modulated to generate features which may explain the low-$\ell$ suppression of power in the CMB temperature anisotropies. Moreover we find that, the ``non-attractor to slow-roll'' evolution retains the large non-Gaussianity generated in the first stage but modified the amplitude.  Meanwhile there are also step features in both the local and equilateral limits of the bispectrum.

The article is organized as follows. In Section~\ref{sec:non-attractor}, we quickly review the non-attractor inflation and the calculation of the primordial curvature perturbation during this stage, which leads to the limited $e$-folding number. In Section~\ref{sec:slow-roll}, we first discuss which type of slow-roll models could be connected to the non-attractor phase, and then study the background evolution with a relaxation stage and the resulting power spectrum. In Section~\ref{sec:nG}, we use the formalism to calculate the bispectrum and see the interesting features caused by the non-attractor initial condition. In Section~\ref{sec:plateau}, we briefly remark the implications on the inflation models with plateau-like potentials. In Section~\ref{sec:discussions}, we give some discussion and conclude this work.

\section{Non-attractor inflation}
\label{sec:non-attractor}

\subsection{Background evolution}

\begin{figure}[tbhp]
\centering
\includegraphics[width=0.5\linewidth]{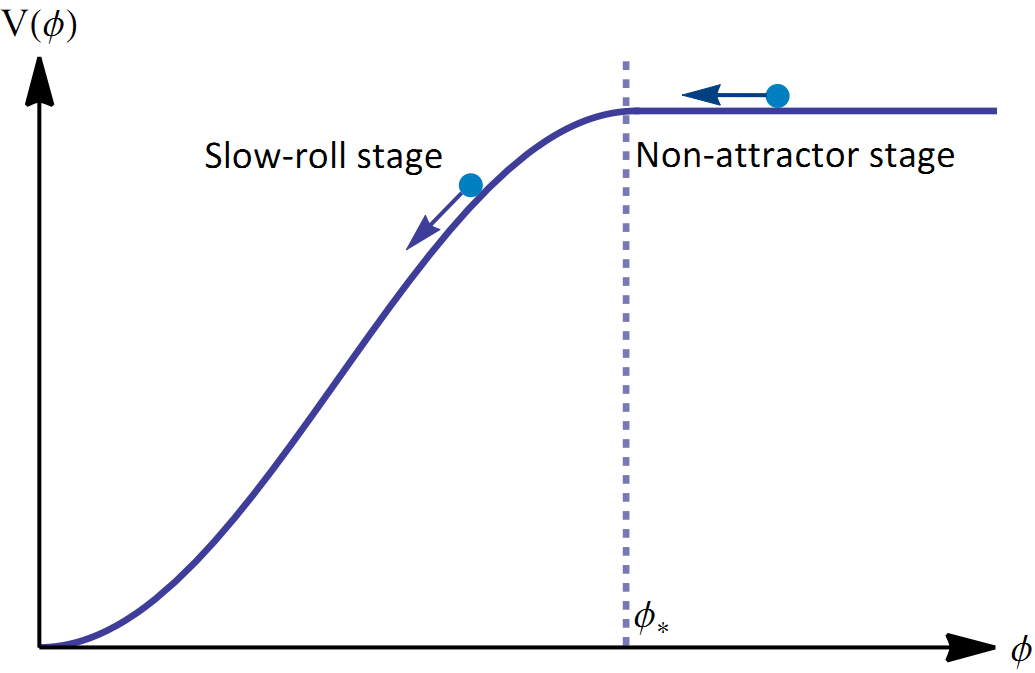}
\caption{An example of a non-attractor potential followed by a slow-roll stage.}
\label{fig:potential}
\end{figure}

\begin{figure}[tbhp]
\centering
\includegraphics[width=0.7\linewidth]{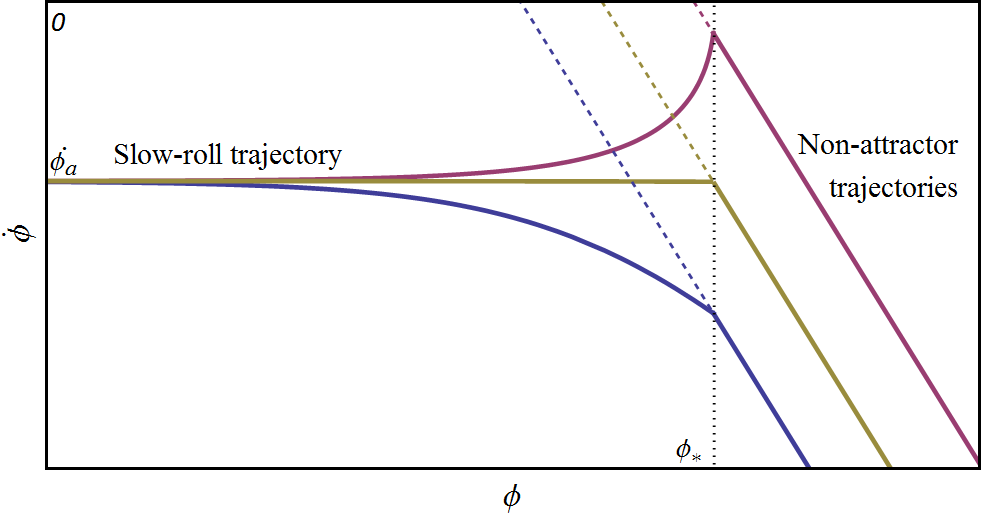}
\caption{The phase space diagram around the transition point. Here we take chaotic inflation $V(\phi)=m^2\phi^2/2$ as an example for the slow-roll trajectory, in which attractor velocity $\dot\phi_a$ is constant. Obviously there are three possible cases around $\phi_*$: $|\dot\phi_a|>|\pi_*|$, $|\dot\phi_a|=|\pi_*|$ and $|\dot\phi_a|<|\pi_*|$.}
\label{fig:3cases}
\end{figure}

For concreteness, we concentrate on the non-attractor model with a constant potential $V(\phi)=V_0$~\cite{Tsamis:2003px,Kinney:2005vj}, as shown in Figure~\ref{fig:potential}. Since the expansion is driven by the potential energy during inflation, we require the contribution of the kinetic energy to be much smaller, i.e. $\dot\phi^2 \ll V_0$. Then the background evolution equations are
\begin{equation}
\ddot{\phi}+3H\dot{\phi}=0
\quad \text{and} \quad
H^2=\frac{1}{3\mpl^2}\(\frac{1}{2}\dot{\phi}^2+V_0\) \approx \frac{1}{3\mpl^2}V_0~.
\end{equation}
Setting the non-attractor stage ends at $t=t_*$ with $\phi(t_*) \equiv \phi_*$ and $\dot{\phi}(t_*) \equiv \pi_*$, we can solve the above two equations exactly with the solutions
\begin{equation}
\label{phit}
\phi(t)=\phi_*+\frac{\pi_*}{3H} \left[ 1- e^{-3H(t-t_*)} \right]
\quad \text{and} \quad
a(t)=a_*e^{H(t-t_*)}~.
\end{equation}
Without losing generality, we normalize $a_*=1$. Then the velocity of the inflaton on the constant potential is described by
\begin{equation}
\label{dphit}
\dot{\phi}(t) = \pi_*e^{-3H(t-t_*)} = \pi_*a^{-3}(t)~,
\end{equation}
while the trajectories of the inflaton in the phase space is given by $\dot\phi+3H\phi = 3H\phi_*+\pi_*$ = constant, as shown in Figure~\ref{fig:3cases}. Note that differing from the slow-roll case in the next section, these trajectories depend on {\em initial values} of both $\phi$ and $\dot\phi$, so there is no unique attractor trajectory.

The $e$-folding number counted from $t$ to $t_*$ is defined as
\begin{equation}
\label{nonattN}
N(t_*,t) \equiv \int_t^{t_*} Hdt = H(t_*-t)~.
\end{equation}
Then using \eqref{phit} and \eqref{dphit}, we obtain $3H(\phi-\phi_*)=\dot\phi(e^{-3N}-1)$, which leads us to the expression of $N$ as~\cite{Namjoo:2012aa,Chen:2013eea}
\begin{equation}
\label{naN}
N(\phi, \dot{\phi}; \phi_*) = \frac{1}{3}\log \[ \frac{\dot\phi}{\dot\phi+3H(\phi-\phi_*)} \]~.
\end{equation}
Note that the $e$-folding number is a function of both $\phi$ and $\dot\phi$. This can be explained by the dependence of inflaton trajectories on $(\phi, \phi_*)$ in the phase space. Thus in this case, $\dot\phi$ is another independent parameter other than $\phi$ that influences the background evolution\footnote{See~\cite{Cai:2015xba} for more discussion on the generalization of the so-called $\delta{N}$ formalism and another case in which the $e$-folding number includes the speed of sound.}.

Introducing the first slow-roll parameter
\begin{equation}
\label{epsilon}
\epsilon \equiv -\frac{\dot H}{H^2} = \frac{1}{2\mpl^2}\frac{\dot\phi^2}{H^2}~,
\end{equation}
during the non-attractor stage it decays very quickly as $\epsilon\propto a^{-6}$. Then the second slow-roll parameter follows directly as
\begin{equation}
\label{eta}
\eta \equiv \frac{\dot\epsilon}{H\epsilon} = -6~,
\end{equation}
which is not a small quantity in the non-attractor case.

\subsection{Power spectrum}

Having worked out the background dynamics, now we turn to consider the comoving curvature perturbation $\calR$. One unique feature for non-attractor inflation is that $\calR$ grows on super-horizon scales. In the previous treatments~\cite{Namjoo:2012aa,Chen:2013eea}, the conventional $\delta N$ formalism is extended based on \eqref{naN}, so that the growing behavior of $\calR$ is included. Here we use another, simpler approach to tackle the evolution of curvature perturbation.

On super-horizon scales, the equation of motion of $\calR$ becomes~\cite{Wang:2013eqj}
\begin{equation}
\ddot\calR + (3+\eta)H\dot\calR = 0~,
\end{equation}
whose general solution takes the following form:
\begin{equation}
\calR = c_1 + c_2\int^{t}dt_1\exp\left\{-\int^{t_1}[3+\eta(t_2)]H(t_2)dt_2\right\}~.
\end{equation}
The first term is constant, while the second one is time-varying. Here we can figure out a criterion for the conservation of $\calR$: if $\eta>-3$, the second term is decaying with time, so $\calR$ remains constant outside the horizon and we can safely calculate the amplitude of curvature perturbation at horizon-exit. This is the case for the conventional slow-roll inflation, where $\eta\simeq0$. If $\eta<-3$ as we get for non-attractor phase, the second term continues to increase and dominates the first one, thus $\calR$ would grow after horizon-exit until $\eta=-3$. For this case, the spectrum should be evaluated when $\eta$ reaches $-3$ so that $\calR$ becomes constant eventually.

As is well known, the comoving curvature perturbation is related to the inflaton fluctuation $\delta\phi$ on flat hypersurfaces as
\begin{equation}
\calR = -\frac{H}{\dot\phi}\delta\phi~.
\end{equation}
Moving to the Fourier space, the corresponding power spectrum is
\begin{equation}
\calP_\calR \equiv \frac{k^3}{2\pi^2} |\calR_k|^2 = \frac{k^3}{2\pi^2} \( \frac{H}{\dot\phi} \)^2 |\delta\phi_k|^2 = \(\frac{H}{2\pi}\)^2 \( \frac{H}{\dot\phi} \)^2~,
\end{equation}
where we have used $\calP_{\delta\phi} = k^3|\delta\phi_k|^2/(2\pi^2) = [H/(2\pi)]^2$ in a quasi-de Sitter space for a light scalar field. Recall that for the non-attractor inflation, $\calR$ grows after horizon crossing. The spectrum above should be evaluated at the end of this stage, thus we obtain an exactly scale-invariant spectrum at $t_*$:
\begin{equation}
\label{naspec}
\calP_\calR^\text{na}(t_*) = \(\frac{H}{2\pi}\)^2 \frac{H^2}{\dot\phi^2(t_*)} =\(\frac{H^2}{2\pi{\pi}_*}\)^2
\quad \text{for} \quad
k < k_*~,
\end{equation}
where $k_*$ is the wavenumber whose perturbation mode exits the horizon at the end of the non-attractor stage, i.e. $k_*=a(t_*)H(t_*)$.

With the background above, now we come to check the $e$-folding number during non-attractor inflation. As we have seen, the kinetic energy decreases very quickly as $T = \dot\phi^2/2 \propto a^{-6}$. If we denote the total $e$-folding number of this stage as $N_*$, the kinetic energy at some earlier moment is exponentially larger than that at the end of the attractor stage, $T=T_*e^{6N_*}$. Since the inflaton potential must dominate the kinetic energy during inflation, the largest $T_{\rm max}$ should be smaller than the potential energy. Here we set $T_{\rm max}=\gamma V_0$ and take the numerical factor $\gamma=\calO(0.1)$ for example. Then at the end of the non-attractor stage we have $\pi_*^2=2T_{\rm max}^2/e^{6N_*}=2\gamma V_0/e^{6N_*}$. Substituting this relation into \eqref{naspec} and using $3\mpl^2H^2=V_0$, finally the spectrum can be expressed as
\begin{equation}
\calP_\calR^\text{na} = \frac{e^{6N_*}V_0}{72\gamma\pi^2\mpl^4}~.
\end{equation}
Since the CMB observations yield $\calP_\calR \simeq 2.5\times10^{-9}$~\cite{Ade:2015lrj}, for $N_*\sim4$ we could get this amplitude by lowering the energy scale of inflation to $V_0\sim10^{14}$GeV, but for $N_*\sim60$ there seems no way to make it work. So the duration of the non-attractor phase must be highly  limited.

Given the constant potential, we also face the graceful exit problem. And the slight red tilt of the primordial spectrum excludes the constant potential all the way. All in all, to achieve sufficient $e$-foldings and to terminate the non-attractor phase leading to the correct spectral index, we need to introduce a slow-roll phase following the non-attractor evolution~\cite{Namjoo:2012aa}.

\section{Slow-roll phase of inflation}
\label{sec:slow-roll}

In the last section we have seen that a slow-roll stage is necessary to complete the non-attractor stage. But not all slow-roll models are capable of being attached to the constant potential. In this section, we first show that, to be consistent with observations, certain conditions must be satisfied for the slow-roll potential. Moreover, before joining the slow-roll trajectory, generally there is a relaxation stage between the non-attractor and slow-roll evolution, which is important for the behavior of the curvature perturbation. Therefore, we first give a detailed discussion of this transition phase.

\subsection{How to add a slow-roll stage?}

We assume the slow-roll potential has the general form $V(\phi)$, which is continuously matched to the constant potential at the transition point $\phi_*$, i.e. $V(\phi_*)=V_0$. Then the background equations are
\begin{equation}
\label{bgeq}
\ddot{\phi}+3H\dot{\phi}+V'(\phi)=0
\quad \text{and} \quad
H^2=\frac{1}{3\mpl^2}\[\frac{1}{2}\dot{\phi}^2+V(\phi)\]~.
\end{equation}
Slow-roll evolution requires $\epsilon\ll1$ and $\eta\ll1$, which means $\dot\phi^2\ll V(\phi)$ and $\ddot\phi\ll3H\dot\phi$ respectively. Then we can find the unique trajectory from the above two equations:
\begin{equation}
\label{srtraj}
\dot\phi_a=-\frac{\mpl}{\sqrt3}\frac{V'}{V^{1/2}}~,
\end{equation}
The unique attractor solution indicates that $\phi$ and $\dot\phi$ here are not independent, instead they are related to each other via the slow-roll trajectory.

For the value of $\dot\phi_a$ at the transition point $\phi_*$, there are three possible cases: $|\dot\phi_a(t_*)|>|\pi_*|$, $|\dot\phi_a(t_*)|=|\pi_*|$ and $|\dot\phi_a(t_*)|<|\pi_*|$, as shown in Figure~\ref{fig:3cases}. Here we give a rough estimation to show that the first case above is strongly disfavored by the CMB observations. As is well-known, the power spectrum for the slow-roll stage reads (see later)
\begin{equation}
\calP_\calR^\text{sr}(k) = \( \frac{H}{2\pi} \)^2 \( \frac{H}{\dot\phi_a(t_k)} \)^2~,
\end{equation}
where $t_k$ is the horizon-crossing time for the mode with wavenumber $k$, i.e. $k=a(t_k)H(t_k)$. Recall that the scale-invariant spectrum generated during the non-attractor stage is given by \eqref{naspec}. Since the potential around the transition is still very flat, we can safely assume the variation of the Hubble constant is negligible. Then for $|\dot\phi_a|>|\pi_*|$ we have $\calP_\calR^\text{na} > \calP_\calR^\text{sr}$. Since $\calP_\calR^\text{na}$ corresponds to the spectrum at small wavenumbers $k<k_*$, this indicates an enhancement on large scales. While the CMB observations favor a spectrum suppressed at small $k$, thus we focus on the cases $|\dot\phi_a|\leq|\pi_*|$ in the following.

As we have argued in the previous section, in order that the non-attractor stage lasts for enough $e$-folding numbers to show its effects, $\pi_*$ should be very close to zero. Specifically we have $\pi_*\sim\mathcal O(10^{-13})$ for $N_*\simeq3$ and $\gamma\simeq0.1$. Then $|\dot\phi_a|\leq|\pi_*|$ results in $\dot\phi_a\simeq0$. From \eqref{srtraj}, we find that $\dot\phi=0$ at the beginning of slow-roll stage is equivalent to $V'(\phi)=0$ at some finite field value $\phi$ around $\phi_*$. This argument leads us to hilltop inflation. Generally we can formulate the potential around the transition point in the following form:
\begin{equation}
\label{eq:SRpotential}
V(\phi)=V_0-\frac{\alpha}{p}(\phi-\phi_*)^p+\mathcal{O}\((\phi-\phi_*)^{p+1}\)~,
\end{equation}
with $p>1$. For the validity of this expansion, we need to focus on the field excursion $|\phi-\phi_*|\ll1$. Since $\dot\phi$ is very small on the hilltop, this condition can be held for sufficient $e$-folding numbers, which are responsible for the CMB observations. Otherwise the higher-order contributions to the potential would be significant, which may help to end inflation and post-inflationary processes like reheating.

\subsection{Relaxation process between two stages}

\begin{figure}[tbhp]
\centering
\includegraphics[width=0.7\linewidth]{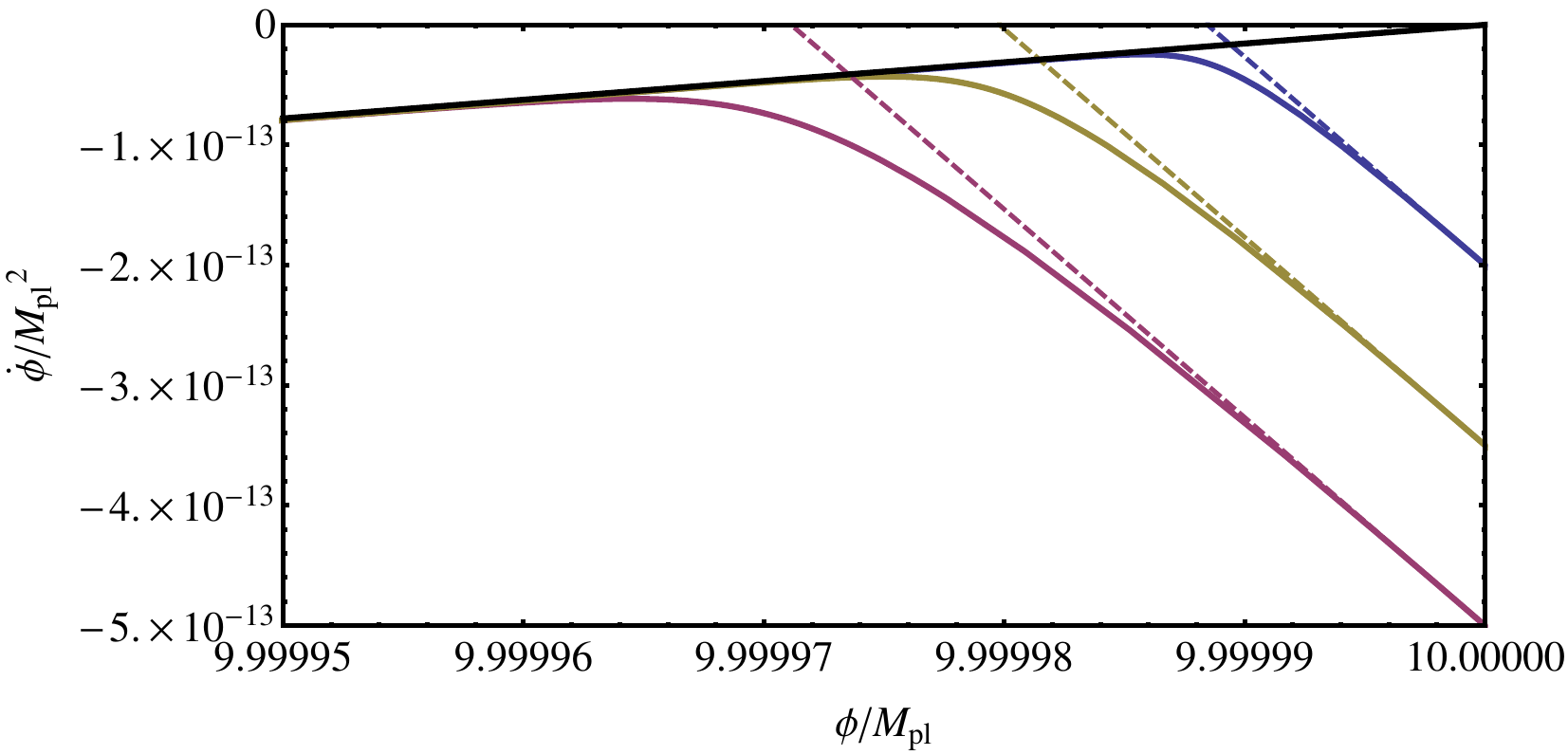}
\caption{The phase space diagram for the relaxation stage. Different solid lines correspond to different initial momentum $|\pi_*|$. The black line is the slow-roll trajectory. The dashed colored lines are the non-attractor trajectories if there is no hilltop potential and the constant one continues. In this figure we take $p=2$, $\phi_*=10\mpl$ and $\alpha=0.3V_0$ as an example.}
\label{fig:relaxation}
\end{figure}

Since at the transition point, generally $\dot\phi_a\neq\pi_*$, there is a relaxation stage during which the inflaton evolves from the non-attractor phase to the slow-roll phase. The background equations during this stage just follow from \eqref{bgeq} and \eqref{eq:SRpotential} as
\begin{equation}
\label{bgrela}
\ddot\phi + 3H\dot\phi - \alpha(\phi-\phi_*)^{p-1} = 0
\quad \text{and} \quad
3H^2\mpl^2 = \frac{1}{2}\dot\phi^2 + V(\phi) \approx V_0~.
\end{equation}
The second equation makes use of the approximation $|\phi-\phi_*|\ll0$ , so the Hubble parameter here is also constant, then we can still use the de Sitter solution $a=a_*e^{H(t-t_*)}$ to describe the background expansion. In the following we shall mainly take $p=2$ as an example, which can be solved analytically. Given $\phi(t_*)=\phi_*$ and $\dot\phi(t_*)=\pi_*$ at the transition point, $\phi(t_e)$ and $\dot\phi(t_e)$ at some later time $t_e$ can be expressed as
\begin{align}
\label{relaxphi}
\phi(t_e) & = \phi_*+\frac{\pi_*}{3sH} \[ a_e^{3(s-1)/2} - a_e^{-3(s+1)/2} \]~,
\\
\label{relaxdphi}
\dot\phi(t_e) & = \frac{\pi_*}{2} \[ \frac{s+1}{s}a_e^{-3(1+s)/2} + \frac{s-1}{s}a_e^{-3(1-s)/2} \]~,
\end{align}
where $a_e = a(t_e)$ and $s \equiv \sqrt{1+4\alpha/(9H^2)}$. Since inflation requires the effective mass is much smaller than the energy scale $\alpha\ll H^2$, we have $s
\simeq1$. Its evolution in the phase space is shown in Figure~\ref{fig:relaxation}. As we can see, the relaxation trajectories deviate from the non-attractor gradually and then join the slow-roll evolution.

\begin{figure}[tbhp]
\centering
\includegraphics[width=0.45\linewidth]{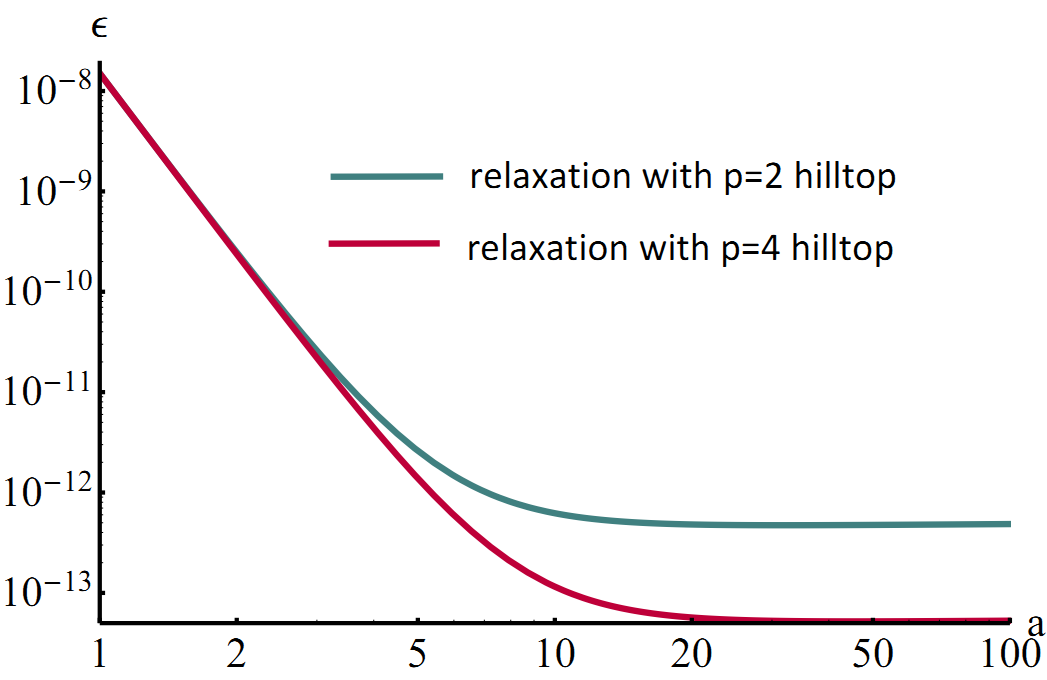}
\hspace{1em}
\includegraphics[width=0.45\linewidth]{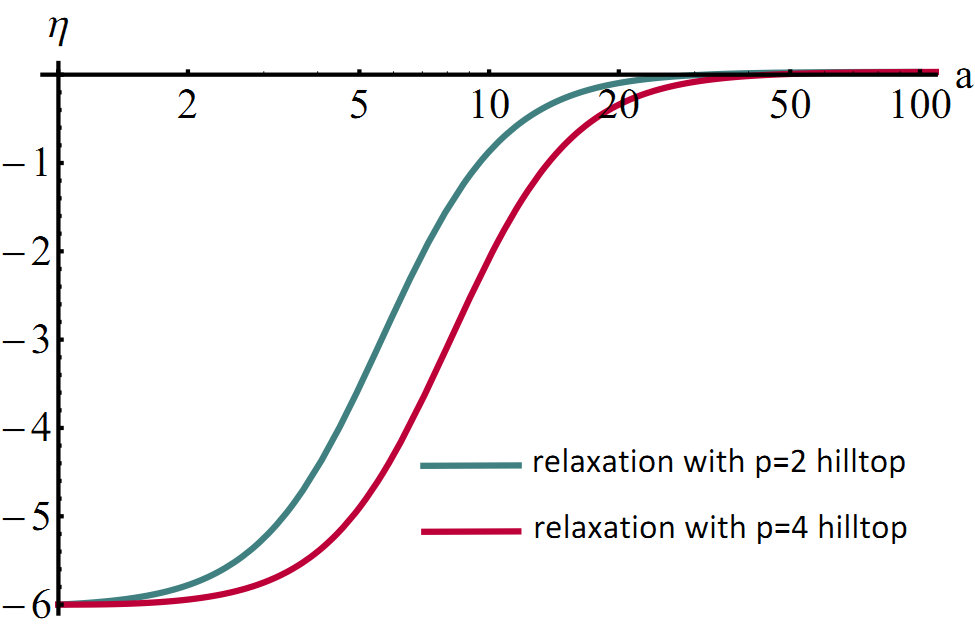}
\caption{The evolution of the two slow-roll parameters $\epsilon$ and $\eta$ with scale factor $a$ during the relaxation stage.}
\label{fig:srpara}
\end{figure}

The slow-roll parameters at time $t_e$ in this stage are given by
\begin{align}
\label{epsilon2}
\epsilon & = \frac{\pi_*^2}{8\mpl^2H^2} \[ \( 1+\frac{1}{s} \)^2a_e^{-3-3s} + \( 1-\frac{1}{s^2} \) a_e^{-3} + \( 1-\frac{1}{s} \)^2 a_e^{-3+3s} \]~,
\\
\label{eta2}
\eta & = 3\frac{a_e^{3s}(s-1)^2-(1+s)^2}{a_e^{3s}(s-1)+(1+s)}~.
\end{align}
Their behavior is shown in Figure~\ref{fig:srpara}. At the beginning of the relaxation stage, $a\rightarrow1$, we have $\epsilon\propto a^{-6}$ and $\eta\simeq-6$, which are still in the non-attractor evolution. But for $a\gg1$, we return to the slow-roll result $\epsilon\to$ constant and $\eta\simeq 2\alpha/(3H^2)$. The $p=4$ case of hilltop can be solved numerically and the corresponding $\epsilon$ and $\eta$ are also shown in Figure~\ref{fig:srpara}.

\subsection{Power spectrum}

\begin{figure}[tbhp]
\centering
\includegraphics[width=0.45\linewidth]{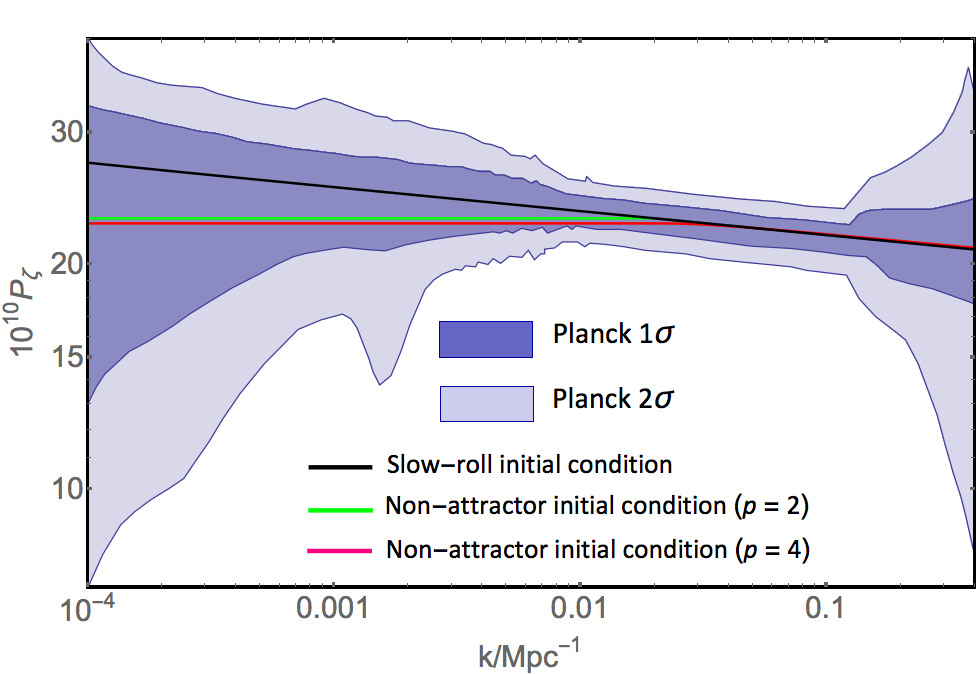}~~\includegraphics[width=0.465\linewidth]{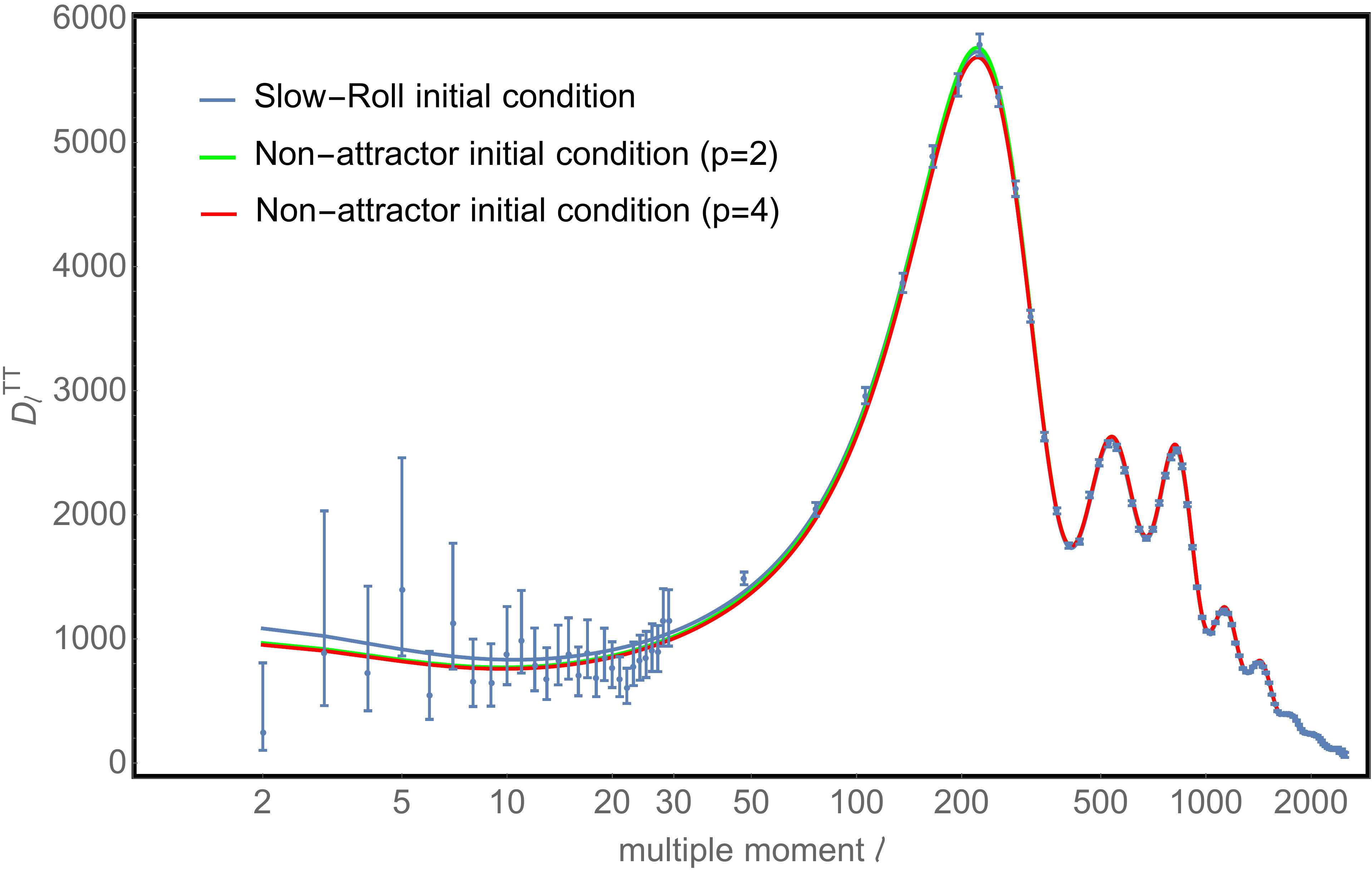}
\caption{The primordial spectrum of the curvature perturbation (left) and the CMB $TT$ spectrum (right) with the constraints of Planck. The spectrum generated during non-attractor stage is scale-invariant, while the one generated during slow-roll stage is red-tilted. The scale-invariant part can be seen as the suppression on large-scales.}
\label{fig:spectrum}
\end{figure}

Following the conventional treatment, we define $z \equiv a\mpl\sqrt{2\epsilon}$ and $u_k \equiv z\calR_k$, and then we find the Mukhanov-Sasaki equation
\begin{equation}
u_k'' + \( k^2 -\frac{z''}{z} \) u_k = 0~,
\end{equation}
where $z''/z$ is exactly written in terms of the slow-roll parameters as
\begin{equation}
\frac{z''}{z} = a^2H^2 \( 2+\frac{3}{2}\eta-\epsilon-\frac{1}{2}\epsilon\eta+\frac{1}{4}\eta^2+\frac{1}{2}\delta\eta \)~.
\end{equation}
Here $\delta \equiv \dot\eta/(H\eta)$ is another ``slow-roll'' parameter. As we have argued, around the transition point $\phi_*$ the potential remains flat enough, so the expanding background still can be seen as an exact de Sitter universe and $a^2H^2={1}/{\tau^2}$. Then using the standard treatment, we set
\begin{equation}
\label{nu}
\nu^2 \equiv \frac{9}{4} +\frac{3}{2}\eta-\epsilon-\frac{1}{2}\epsilon\eta+\frac{1}{4}\eta^2+\frac{1}{2}\delta\eta~.
\end{equation}
In the non-attractor stage, $\nu=3/2$. In the relaxation stage, although the slow-roll parameters \eqref{epsilon2} and \eqref{eta2} undergo drastic variations, we have $\nu=3s/2$ if the terms with $\epsilon$ are ignored. Thus for a positive and constant $\nu$, the mode function with initial Bunch-Davis vacuum can be solved as
\begin{equation} 
\label{modef}
u_k\propto \sqrt{-\tau} H_\nu^{(1)}(-k\tau)~,
\end{equation}
which scales as $\propto k^{-\nu}$ outside the horizon $-k\tau\ll1$. Then the power spectrum scales directly $\calP_\calR \propto k^3|u_k|^2 \propto k^{3-2\nu}$, which results in the spectral index
\begin{equation}
\label{ns}
n_\calR-1 \equiv \frac{d\log\calP_\calR}{d\log k} = 3-2\nu~.
\end{equation}
Since it should be calculated at horizon exit $k=a(t_k)H$, we obtain a spectral index with slight $k$-dependence. During the non-attractor stage $n_\calR-1=0$ as we have seen previously. The spectral index generated in the following stage is slightly red $n_\calR-1 = 3(1-s) \simeq -2\alpha/(3H^2)$, which is constant\footnote{A similar calculation of the spectrum is performed in Ref.~\cite{Tzirakis:2007bf}, where the authors considered the non-slow-roll evolution on a hilltop potential, which is related to the relaxation stage in our case.}. The whole spectrum is shown in Figure~\ref{fig:spectrum}. The behavior of the spectrum at small $k$ may well account for the large scale suppression in the CMB $TT$ spectrum. The numerical results of $p=4$ is also shown in Figure~\ref{fig:spectrum}.

\section{Features in the primordial non-Gaussianity}
\label{sec:nG}

Now we study the bispectrum of the primordial perturbation. The non-attractor stage is well known to violate of non-Gaussianity consistency relation in the framework of single-field models, which is verified via both the $\delta N$ formalism and in-in calculation~\cite{Namjoo:2012aa,Chen:2013eea} (see however~\cite{Mooij:2015yka,Romano:2016gop}). So it is expected that the non-attractor initial conditions here would yield non-trivial features in the primordial bispectrum.
In this section, we give the first principle calculation of non-Gaussianity using the in-in formalism.

We begin with the cubic action\footnote{In this section, we set $\mpl=1$ and $p=2$ for simplicity.}~\cite{Maldacena:2002vr}:
\begin{align}
S_3 & = \int dtd^3x \bigg[
a^3\epsilon^2 \calR\dot{\calR}^2+a\epsilon^2 \calR(\nabla\calR)^2-
2a \epsilon\dot{\calR}(\partial^i\calR)(\partial_i\chi)
\nonumber
\\
& \hspace{3em} +
\frac{a^3\epsilon}{2}\dot\eta \calR^2\dot{\calR}
+\frac{\epsilon}{2a}(\partial^i\calR)(\partial_i\chi) \Delta\chi + \frac{\epsilon}{4a}(\Delta\calR)(\nabla\chi)^2
 + 2f(\calR)\frac{\delta L}{\delta \calR} \bigg|_1 \bigg] ~,
\label{action_cubic}
\end{align}
where $\Delta \equiv \delta^{ij}\partial_i\partial_j$ and
\begin{align}
\chi & = a^2 \epsilon \Delta^{-1} \dot \calR ~, 
\\
\frac{\delta L}{\delta\calR}\bigg|_1 & = a \Delta
\left( \frac{d\chi}{dt} + H\chi - \epsilon\calR \right) ~, 
\\
f(\calR) & = \frac{\eta}{4}\calR^2 + \frac{1}{H}\calR\dot\calR + \cdots
~.
\label{redefinition}
\end{align}
Using a field redefinition
\begin{equation} 
\label{FR}
\calR = \calR_n +\frac{\eta}{4}\calR_n^2 + \frac{1}{H}\calR_n \dot\calR_n~,
\end{equation}
we can remove the last term in the action \eqref{action_cubic}. As $\epsilon$ is always small while $\eta$ is not, so the terms proportional to $\epsilon^2$ result in slow-roll suppressed non-Gaussianty as in the conventional case. The leading contributions to the bispectrum come from the terms with $\eta$.

\subsection{The fall of the field redefinition}

First we study the contribution from the field redefinition \eqref{FR}, which is equivalent to the contribution obtained by the $\delta{N}$ formalism~\cite{Namjoo:2012aa} and leads to large non-Gaussianity in the non-attractor stage. Here we further explore its behavior during the relaxation stage.

\begin{figure}[tbhp]
\centering
\includegraphics[width=0.6\linewidth]{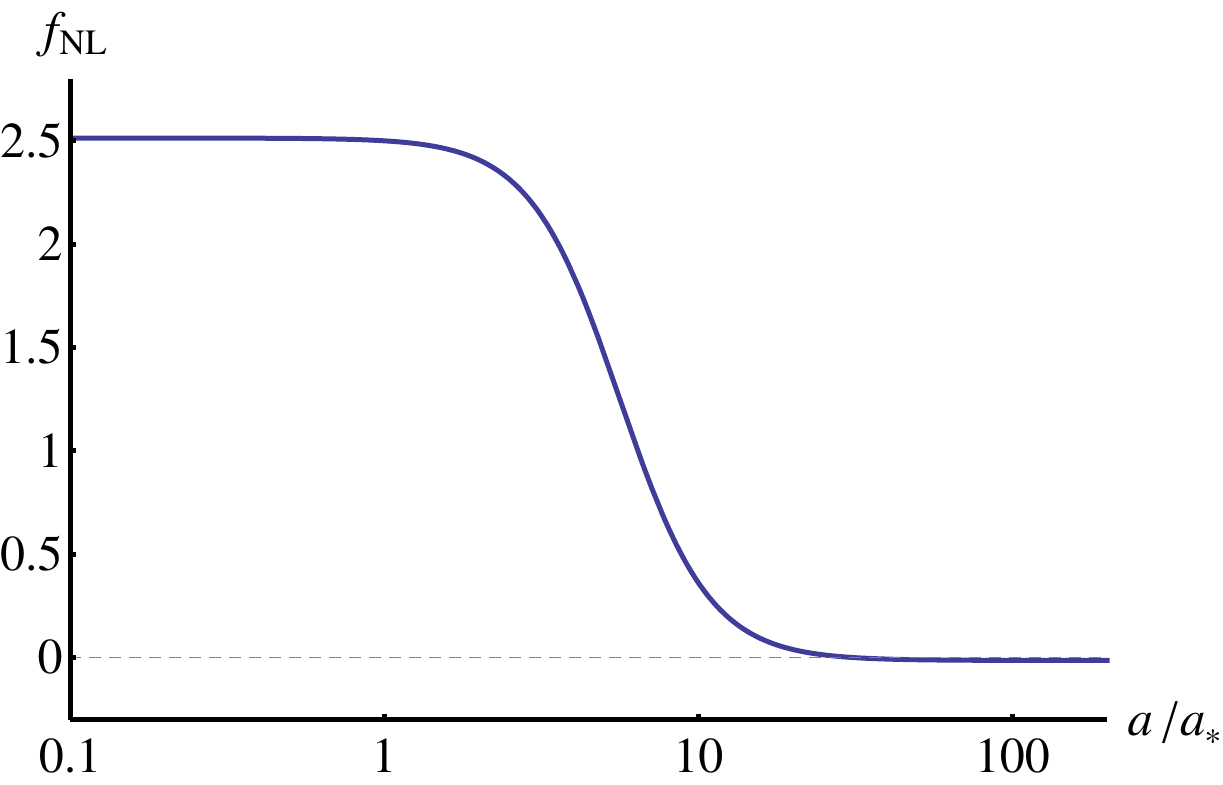}
\caption{The evolution of local $\fnl$ from the field redefinition.}
\label{fig:fnlFR}
\end{figure}

Using the solution (\ref{modef}), we obtain the following expression of curvature perturbation:
\begin{equation}
\calR_k=\frac{u_k}{z}\propto \frac{H(-\tau)^{(3/2)}}{\sqrt{\epsilon}}H^{(1)}_\nu(-k\tau)~,
\end{equation}
then using the slow-roll parameters \eqref{epsilon} and \eqref{eta} as well as \eqref{ns}, we have
\begin{equation}
\dot\calR_k=\frac{1}{2}\[(1-n_\calR)-(\eta+2\epsilon)\]H\calR_k + \calO(k\tau)~.
\end{equation}
This relation holds true for the whole stages of inflation. For instance, in the slow-roll stage, $1-n_\calR=\eta+2\epsilon$ thus for super-horizon modes $\dot\calR_k \simeq 0$, which corresponds to the conservation of $\calR$. In the non-attractor stage, $1-n_\calR=0$ and the curvature perturbation grows after horizon exit: $\dot\calR_k = 3H\calR_k$. For the relaxation stage, $1-n_\calR = 3(s-1) \leq \eta$ thus $\dot\calR \simeq -\eta H\calR_k/2$. Eventually the field redefinition can be expressed as
\begin{equation}
\label{FR2}
\calR_{\bm k} = \calR^n_{\bm k} + \frac{1}{2} \[(1-n_\calR)-\frac{\eta}{2}-2\epsilon\] \int \frac{dp^3}{(2\pi)^3} \calR^n_{\bm{k}-\bm{p}}\calR^n_{\bm p}~.
\end{equation}
Thus the bispectrum becomes
\begin{equation}
\label{frbispec}
\langle\calR_{\bm{k}_1}\calR_{\bm{k}_2}\calR_{\bm{k}_3}\rangle = (2\pi)^3\delta^3(\bm{k}_1+\bm{k}_2+\bm{k}_3) \frac{1}{4}\[(1-n_\calR)-\frac{\eta}{2}-2\epsilon\] \( \frac{H^2}{2\epsilon k_1^3}\frac{H^2}{2\epsilon k_2^3} +\rm{2~perm} \)~.
\end{equation}
Taking the squeezed limit $k_1=k_2\equiv k\gg k_3$ and comparing with the definition of local $\fnl$
\begin{equation}
\label{fnl}
\lim_{k_3/k\rightarrow0}\langle\calR_{\bm{k}_1}\calR_{\bm{k}_2}\calR_{\bm{k}_3}\rangle =  (2\pi)^3\delta^3(\bm{k}_1+\bm{k}_2+\bm{k}_3) \frac{(2\pi)^4\calP_\calR^2}{k^3k_3^3} \frac{3}{5}\fnl~,
\end{equation}
we find
\begin{equation}
\label{fnlFR2}
\fnl = \frac{5}{6}\[(1-n_\calR)-\frac{\eta}{2}-2\epsilon\]~.
\end{equation}
Note that both $1-n_\calR$ and $\epsilon$ are small, while $\eta$ runs from $-6$ to a small value $3(s-1)$ continuously. Thus during the non-attractor stage $\fnl$ is $5/2$, but once inflation enters the slow-roll stage, it is suppressed as shown in Figure~\ref{fig:fnlFR}. In summary, although the field redefinition contributes a large local non-Gaussianity during the initial non-attractor stage, it will not survive in the following slow-roll evolution.

Finally we would like to mention that, in the relaxation stage we can also extend $\delta N$ formalism to consider the ($\phi, \dot\phi$) dependence of the $e$-folding number, which yields the same result with the field redefinition contribution.

\subsection{The rise of the interaction term}

Now we consider the contribution from the interaction Lagrangian. As argued, because of the smallness of $\epsilon$, only one term remains at the leading order of the cubic action:
\begin{equation}
\label{interact}
S_3 \supset \int dtd^3x \frac{a^3\epsilon}{2}\dot\eta \calR^2\dot{\calR}~.
\end{equation}
The intrinsic bispectrum from this term evaluated at $\tau$, which we eventually take as the end of inflation $\tau=0$, is given by
\begin{align}
\left\langle \calR_{\bm{k}_1}(\tau)\calR_{\bm{k}_2}(\tau)\calR_{\bm{k}_3}(\tau) \right\rangle = -2 \Im \Bigg[ & \int d\tau'\prod_{i} \frac{d^3p_i}{(2\pi)^3}\frac{a^2(\tau')\epsilon(\tau')\eta'{\tau'}}{2}(2\pi)^3 \delta^3(\bm{p}_1+\bm{p}_2+\bm{p}_3) 
\nonumber\\
& \times \calR_{\bm{k}_1}(\tau)\calR_{\bm{k}_2}(\tau)\calR_{\bm{k}_3}(\tau) \calR_{\bm{p}_1}^*(\tau')\calR^*_{\bm{p}_2}(\tau')\calR_{\bm{p}_3}^{\prime*} (\tau') \Bigg]~.
\end{align}
Note here $\calR_{\bm p}$ from the interaction Hamiltonian depends on $\tau'$, and changes rapidly before the slow-roll stage; while $\calR_{\bm k}$ does not. With this consideration, the bispectrum above is expressed as
\begin{align}
\label{bispectr}
\langle\calR_{\bm{k}_1}\calR_{\bm{k}_2}\calR_{\bm{k}_3}\rangle & =
(2\pi)^3\delta^3(\bm{k}_1+\bm{k}_2+\bm{k}_3) \frac{(2\pi)^4\calP_\calR^2}{8k_1^3k_2^3k_3^3}
\nonumber\\
& \quad \times
\Im \left\{ \int_{-\infty(1+i\epsilon)}^0 d\tau' \frac{\eta'}{\sqrt{\epsilon/\epsilon_0}} \frac{e^{-i(k_1+k_2+k_3)\tau'}}{\tau'}
\left[(1+ik_1\tau')(1+ik_2\tau')k_3^2 + \text{2 perm} \right] \right\}~,
\end{align}
where $\epsilon_0\simeq\epsilon(0)$ is the nearly constant slow-roll parameter after the transition. Taking the squeezed limit and comparing with \eqref{fnl}, we get the local $\fnl$
\begin{equation}
\label{localfnl}
\fnl = -\frac{5}{12} \Im \left[ \int_{-\infty(1+i\epsilon)}^0 d\tau'\frac{\eta'}{\sqrt{\epsilon/\epsilon_0}}\frac{e^{-2ik\tau'}}{k\tau'}(1+ik\tau') \right]~.
\end{equation}
Performing the above integration is non-trivial, since the full expressions of $\epsilon$ and $\eta$ in \eqref{epsilon2} and \eqref{eta2} are complicated for analytical calcualtion. One way to simplify the problem is to treat $\dot\eta$ as a delta function~\cite{Namjoo:2012aa}. According to the evolution of $\eta$ during the relaxation stage, we take a sudden transition at $\tau_*$ as $\eta=-6[1-\theta(\tau-\tau_*)]$. In this case, we can easily perform the integration in \eqref{localfnl} and find
\begin{equation}
\fnl=-\frac{5}{2} \left[ \cos(2k\tau_*) - \frac{\sin(2k\tau_*)}{k\tau_*} \right]~.
\end{equation}
For the non-attractor modes that exit the horizon before the transition, simply $\fnl=5/2$. So the contribution from the interaction term \eqref{interact} seems to replace the one from field redefinition. Although the latter one vanishes in the following stage, the final bispectrum remains nonzero. But for the modes during the subsequent slow-roll stage, $\fnl=-5\cos(2k\tau_*)/2$. This non-vanishing oscillations in $\fnl$ can be found when the change in $\eta$ or any other slow-roll parameters are approximated to be sudden~\cite{Arroja:2012ae,Gong:2014spa}. In reality, however, the change is smooth and the non-vanishing oscillations are absent.

\begin{figure}[tbhp]
\centering
\includegraphics[width=0.6\linewidth]{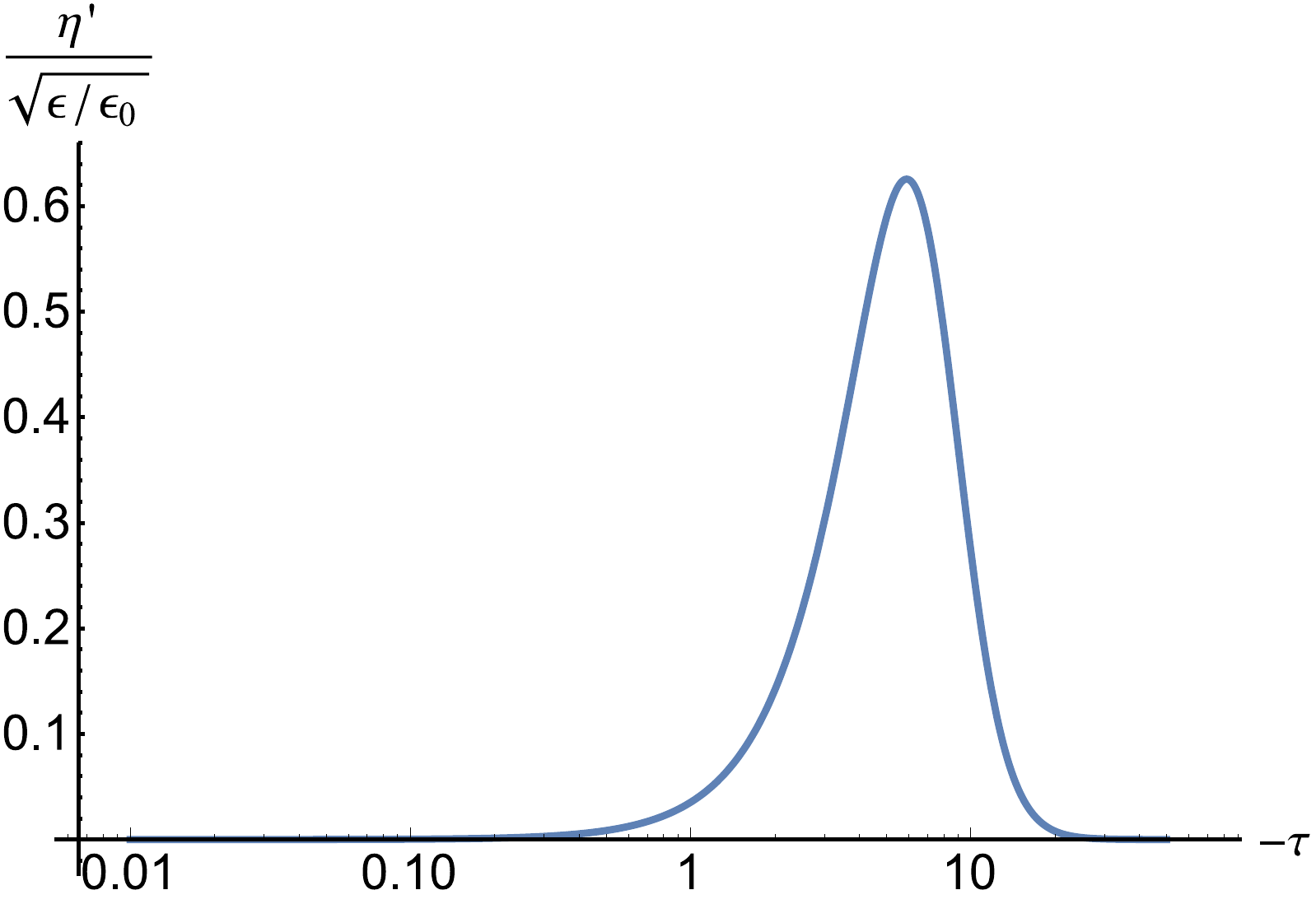}
\caption{The evolution of $\eta'/\sqrt{\epsilon/\epsilon_0}$ with conformal time after the non-attractor stage. Here we set $\tau_*=-50$.}
\label{fig:window}
\end{figure}

\begin{figure}[tbhp]
\centering
\includegraphics[width=0.6\linewidth]{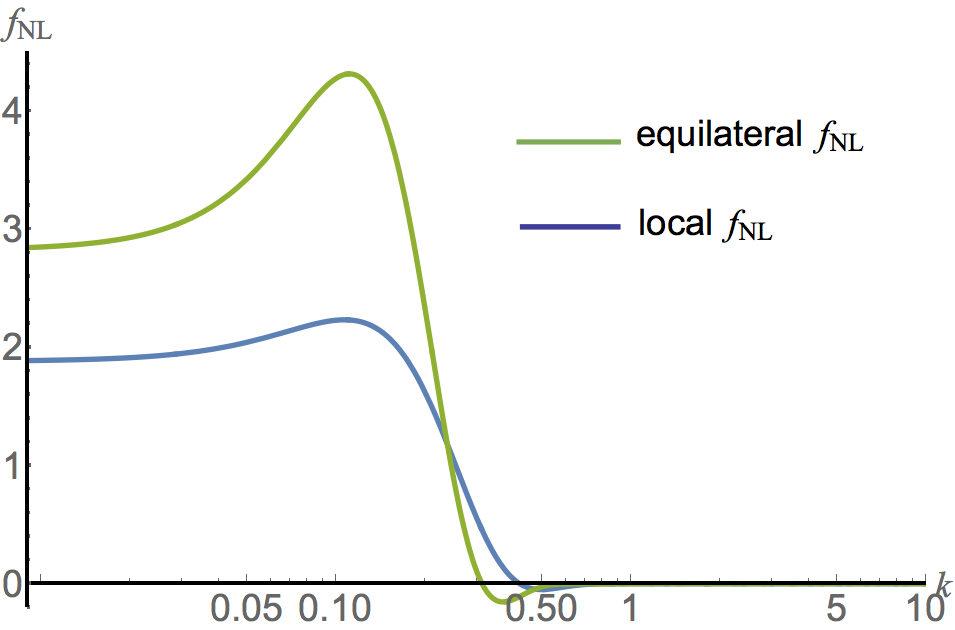}
\caption{Features in the $\fnl$ from interaction term \eqref{interact} for two configurations.}
\label{fig:fnlstep}
\end{figure}

In \eqref{localfnl}, $\eta'/\sqrt{\epsilon/\epsilon_0}$ can be regarded as a window function which has a quite wide range, as shown in Figure~\ref{fig:window}. During both the non-attractor stage and the slow-rollstage. $\eta'$ is negligible. Thus only the relaxation stage matters. Integrating \eqref{localfnl} numerically, the $k$-dependent result is shown in Figure~\ref{fig:fnlstep}. Obviously there are small oscillations around the transition, but it decays as $k$ grows. Thus here we have a step feature of the local non-Gaussianity, while the $\fnl$ generated during the slow-roll phase still remains negligible.

Note that for the non-attractor modes, the local $\fnl$ is around $2$ instead of $5/2$, which differs from the result using the delta function parameterization. The discrepancy is mainly caused by the integration of the window functions: it is $6$ for the delta function approximation of $\eta'$, while $\epsilon$ does not contribute since $ \epsilon(\tau_*)\simeq\epsilon_0$; but in reality we also need to consider the variation of $\epsilon$ during the relaxation stage. Although $\eta$ still runs from $-6$ to $0$, the integration of $\eta'/\sqrt{\epsilon/\epsilon_0}$ deviates from $6$. Thus the non-Gaussianity generated during the non-attractor phase ($\fnl=5/2$) is modified to other nonzero values.

We can further explore other shapes of non-Gaussianity. Here we take the equilateral one as another example. In the equilateral limit $k_1=k_2=k_3=k>k_*$, so the equilateral $\fnl$ is
\begin{equation}
\fnl = -\frac{5}{8} \Im \left[ \int_{-\infty(1+i\epsilon)}^0 d\tau'\frac{\eta'}{\sqrt{\epsilon/\epsilon_0}}\frac{e^{-3ik\tau'}}{k\tau'}(1+ik\tau')^2 \right]~.
\end{equation}
Performing the integration numerically, we obtain the scale-dependent equilateral $\fnl$ as shown in Figure~\ref{fig:fnlstep}, which exhibits similar step features across the transition.

In summary, although the field redefinition contribution to non-Gaussianity does not survive the transition process, the interaction term \eqref{interact} yields non-vanishing local and equilateral $\fnl$ for the non-attractor phase, and they decay to zero for the slow-roll stage. Therefore, even though the initial conditions could result in large non-Gaussianity, it will not last and  the consistency relation $\fnl^{\rm local}=5(1-n_\calR)/12$ is still recovered in the slow-roll stage.

Finally we mention that the calculation above, which divides the contributions to the bispectrum into the one from field redefinition and from interaction terms, can be done without $\dot\eta$. After integrations by part in the cubic action \eqref{action_cubic}, the $\eta$ terms can be written as~\cite{Burrage:2011hd}
\begin{equation}
S_3 \supset\int dtd^3x\[ -a^3\eta\epsilon \calR\dot{\calR}^2-\frac{\epsilon\eta }{2}a\calR^2\Delta\calR\]~.
\end{equation}
Starting from this form, we can find the same result.

\section{Implications for inflation with plateau-like potentials}
\label{sec:plateau}

Using the potential slow-roll parameters
\begin{equation}
\epsilon_V \equiv \frac{\mpl^2}{2}\(\frac{V'}{V}\)^2
\quad \text{and} \quad
\eta_V \equiv \mpl^2\frac{V''}{V}~,
\end{equation}
the scalar spectral index and the tensor-to-scalar ratio of single-field slow-roll inflation can be written as
\begin{equation}
n_\calR-1=-6\epsilon_V+2\eta_V~
\quad \text{and} \quad
r=16\epsilon_V~.
\end{equation}
The observations from the CMB $n_\calR=0.968$~\cite{Ade:2015lrj} and  $r<0.07$~\cite{Array:2015xqh} indicate $\epsilon_V\ll\eta_V$ is likely, favoring a quite flat potential at the beginning of inflation.

\begin{figure}[tbhp]
\centering
\includegraphics[width=0.4\linewidth]{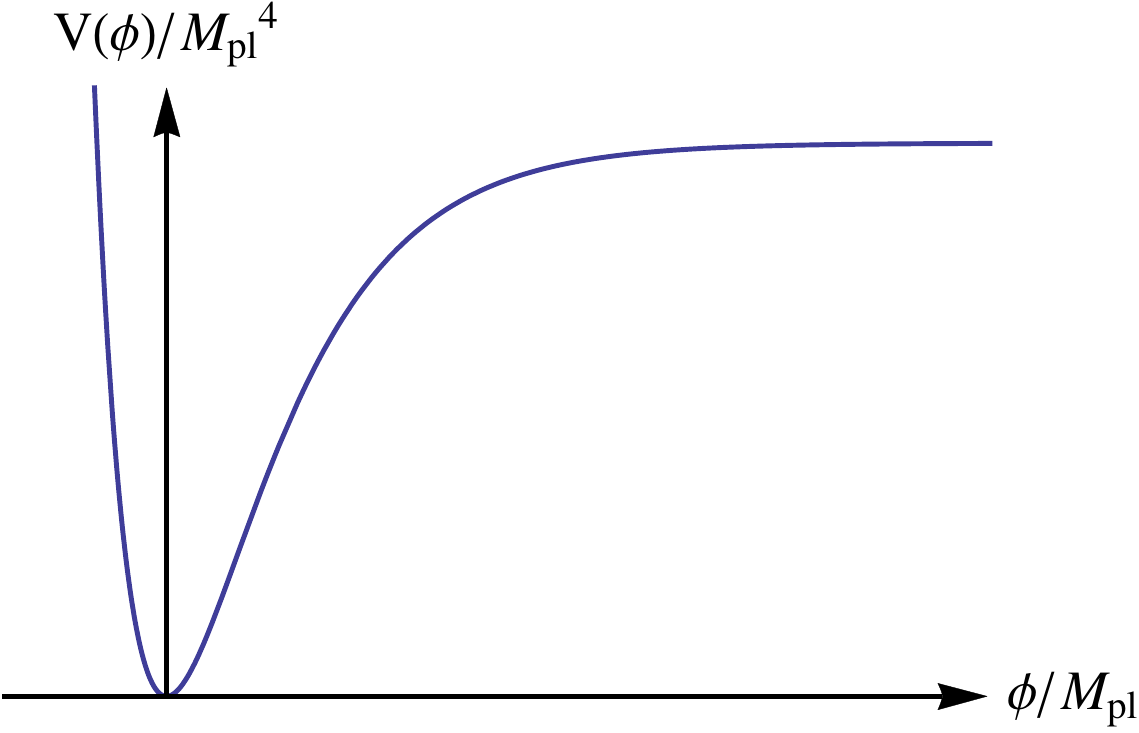}\includegraphics[width=0.4\linewidth]{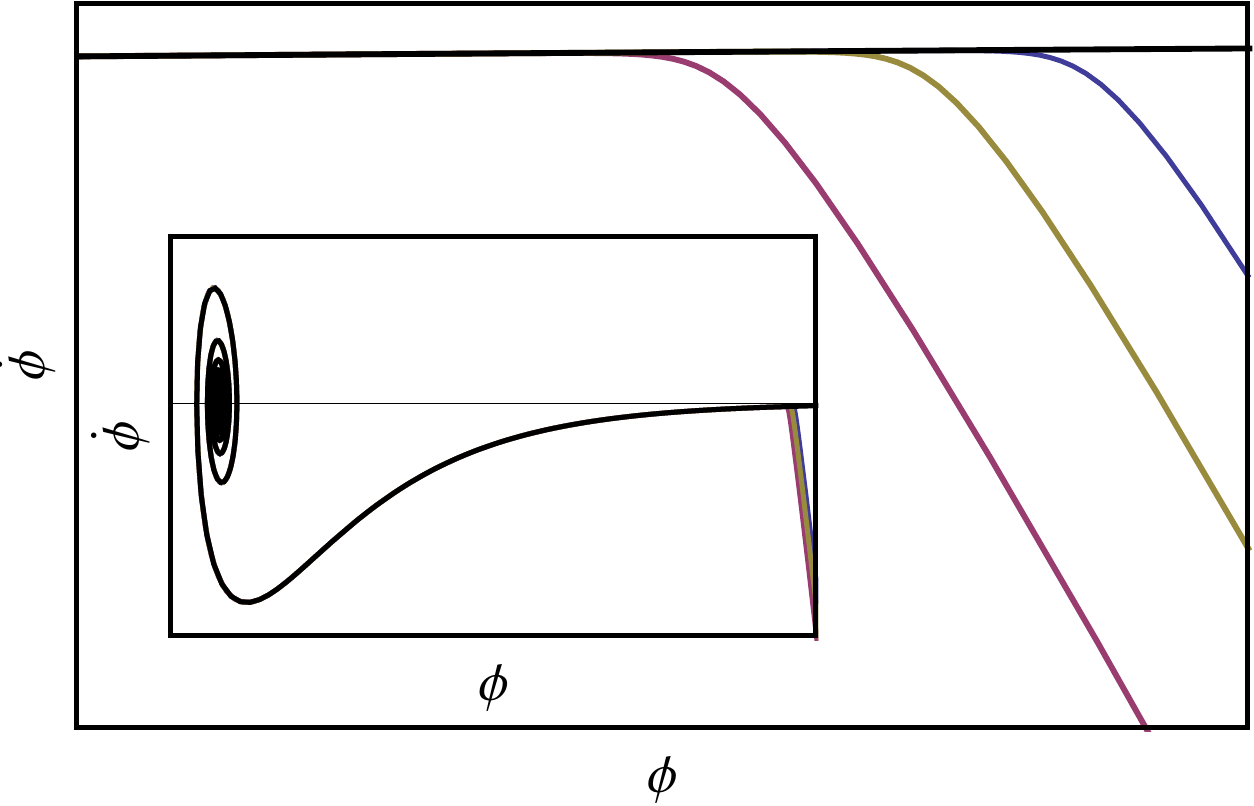}
\caption{Left panel: the potential of $R^2$ inflation; Right panel: trajectories in the phase space diagram. Here the black line is the slow-roll trajectory and the colored lines have non-slow-roll initial conditions. The inner inset shows the complete trajectories while the main plot focuses on the "non-attractor to slow-roll" transition.}
\label{fig:R2D2}
\end{figure}

Comparing the $R^2$ inflation model~\cite{Starobinsky:1980te} as an example for the model with a very flat potential with our phenomenological study, we find many similarities. The effective potential and the slow-roll trajectories in the phase space for the $R^2$ model is shown in Figure~\ref{fig:R2D2}. The phase space diagram shows that if the initial $\dot\phi$ deviates from the slow-roll attractor, we would have a non-attractor stage first before inflaton joins the slow-roll evolution, just like the picture we have discussed in this article. Therefore for models with a ``plateau" potential, if we relax the initial slow-roll attractor conditions, we may explain the low-$\ell$ suppression in the CMB temperature anisotropies and obtain sharp features in $\fnl$.

Before closing this section, we remark that during the non-attractor stage eternal inflation can be evaded. As shown in Figure~\ref{fig:R2D2}, for the slow-roll trajectory $\dot\phi$ approaches zero for larger field values. Thus naively thinking, there must exist a large field value for which the classical field excursion in one Hubble time $\Delta\phi \simeq \dot\phi\Delta{t} \simeq \dot\phi/H$ is smaller than the quantum fluctuation $\delta\phi \simeq H/(2\pi)$ which remains nearly constant. In this region eternal inflation occurs. However, for the non-attractor phase, the CMB normalization of the power spectrum requires $|\Delta\phi| \gg |\delta\phi|$ even at the end of the non-attractor stage. That is, the classical evolution of inflaton always dominates over the quantum fluctuation. Thus the non-attractor initial conditions can help alleviate the problem of eternal inflation for a very flat potential.

\section{Discussions}
\label{sec:discussions}

There are two interesting directions regarding the non-attractor phase during inflation.
First, non-attractor inflation models have aroused a lot theoretical concerns, since in this case the curvature perturbation grows after horizon-exit and the non-Gaussianity consistency relation is violated. Second, from the observational aspects, the latest data favors a class of inflation models, in which the inflaton rolls down a very flat, plateau-like potential. Motivated by these, we study the observable effects of the non-attractor initial conditions of inflation.

In this work, we first review the non-attractor inflation using a toy model with a canonical scalar field, and find that this model can last for only several $e$-folding numbers thus a second stage of inflation should follow. Then to be consistent with the recent observations, the following slow-roll phase of inflation should be driven by hilltop potentials. This model construction is phenomenologically similar to the plateau-like potentials. Next, the power spectrum is calculated and we find that the non-attractor beginning leads to large-scale suppression in the CMB $TT$ spectrum, favored by recent observations. We also work out the bispectrum, and the result shows that large non-Gaussianity generated in the non-attractor stage will not survive the following evolution, but the transition process from non-attractor to slow-roll leads to step features in both the squeezed and equilateral limits of the bispectrum.
These features from the non-attractor beginning of inflation will be interesting for future CMB observations.

Finally we remark that, this work only cares for inflation driven by a single canonical field, which is the simplest case for the non-attractor models. Thus for future researches, it will be interesting to further study the non-attractor initial conditions for more general models with k-essence field. Meanwhile we have assumed the background dynamics is driven by the potential energy all the time. However, it is also tempting to abandon this assumption and consider a kinetic-dominated fast-roll stage before inflation. We expect that further relaxing the initial condition of inflation may result in more interesting features.

\subsection*{Acknowledgements}

We are grateful to Razieh Emami, Chunshan Lin, Junyu Liu, Jerome Quintin, Misao Sasaki, Tao-Tao Qiu, Kai Wang and Yi Wang for helpful discussions. We also thank KITPC for hospitality during the workshop ``Early Universe, Cosmology and Fundamental Physics", where this work was initiated.
YFC, DGW and ZW are supported in part by the Chinese National Youth Thousand Talents Program and by the USTC start-up funding (Grant No.~KY2030000049) and by the National Natural Science Foundation of China (Grant No.~11421303).
JG acknowledges support from the Korea Ministry of Education, Science and Technology, Gyeongsangbuk-Do and Pohang City for Independent Junior Research Groups at the Asia Pacific Center for Theoretical Physics. JG is also supported in part by a Starting Grant through the Basic Science Research Program of the National Research Foundation of Korea (2013R1A1A1006701) and by a TJ Park Science Fellowship of POSCO TJ Park Foundation.

\bibliographystyle{JHEP}%
\bibliography{bibfile}

\end{document}